\documentclass[journal,twocolumn]{IEEEtran}
\usepackage{graphicx,times,amsmath,amssymb,cite,subfigure,stfloats,booktabs,url,multirow,afterpage,float,threeparttable,makecell}
\usepackage[scheme=plain]{ctex}
\allowdisplaybreaks

\makeatletter

\newcommand{\Rmnum}[1]{\expandafter\@slowromancap\romannumeral #1@}
\makeatother

\begin{document}
\title{Spiking Neural Network for Intra-cortical Brain Signal Decoding}

\author{Song Yang, Haotian Fu, Herui Zhang, Peng Zhang, Wei Li and Dongrui Wu,~\IEEEmembership{Fellow, IEEE}
\thanks{S. Yang, H. Fu, H. Zhang, W. Li and D. Wu are with the Key Laboratory of the Ministry of Education for Image Processing and Intelligent Control, School of Artificial Intelligence and Automation, Huazhong University of Science and Technology, Wuhan 430074, China. They are also with Hubei Key Laboratory of Brain-inspired Intelligent Systems, Huazhong University of Science and Technology, Wuhan, 430074, China.}
\thanks{P. Zhang is with Department of Biomedical Engineering, College of Life Science and Technology, Huazhong University of Science and Technology, Wuhan 430074, China.}
\thanks{W. Li and D. Wu are the corresponding authors. Email: liwei0828@hust.edu.cn, drwu09@gmail.com.}}

\maketitle

\begin{abstract}
Decoding brain signals accurately and efficiently is crucial for intra-cortical brain-computer interfaces. Traditional decoding approaches based on neural activity vector features suffer from low accuracy, whereas deep learning based approaches have high computational cost. To improve both the decoding accuracy and efficiency, this paper proposes a spiking neural network (SNN) for effective and energy-efficient intra-cortical brain signal decoding. We also propose a feature fusion approach, which integrates the manually extracted neural activity vector features with those extracted by a deep neural network, to further improve the decoding accuracy. Experiments in decoding motor-related intra-cortical brain signals of two rhesus macaques demonstrated that our SNN model achieved higher accuracy than traditional artificial neural networks; more importantly, it was tens or hundreds of times more efficient. The SNN model is very suitable for high precision and low power applications like intra-cortical brain-computer interfaces.
\end{abstract}


\IEEEpeerreviewmaketitle

\section{Introduction} \label{sect:Introduction}

\begin{figure}[htpb] \centering
	\includegraphics[width=.45\textwidth,clip]{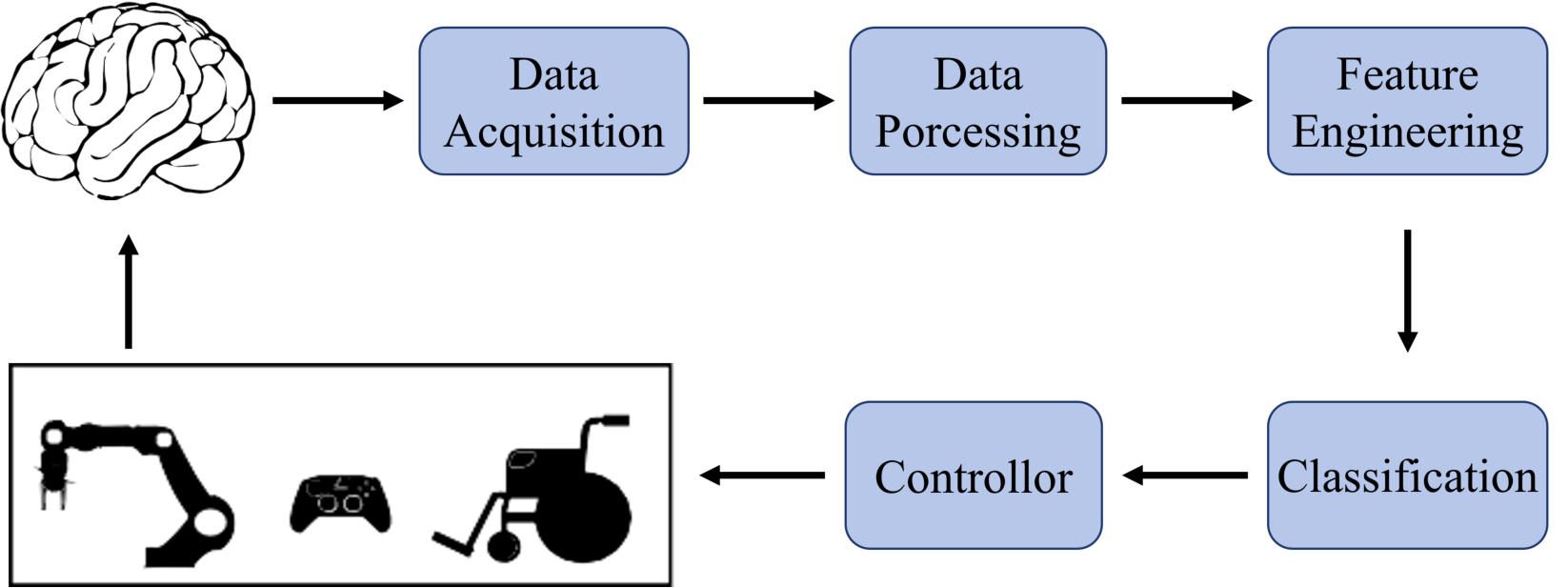}
	\caption{A closed-loop BCI system.} 	\label{fig1}
\end{figure}

Brain-computer interfaces (BCIs) \cite{wolpaw2002brain}, as shown in Fig.~\ref{fig1}, establish a direct communication pathway between the brain and external devices. They can be categorized into invasive BCIs, whose electrodes are placed inside the skull, and non-invasive ones, which do not need surgery.

Non-invasive BCIs can be used for text input \cite{Nakanishi2020}, neuro-rehabilitation \cite{Mane2020}, device control \cite{Leeb2007}, etc. Despite their convenience, the input signals of non-invasive BCIs have low spatial resolution and limited information, and hence may not be suitable for sophisticated applications like accurate speech decoding \cite{Moses2021} and restoring walking abilities of paralyzed patients \cite{Lorach2023}.

Intra-cortical brain-computer interfaces (iBCIs), a type of invasive BCIs that implant electrodes into the cortex, can record much higher quality brain signals, enabling more precise neural signal decoding and more sophisticated BCI applications. iBCIs have demonstrated remarkable capabilities in motor movement interpretation and device control, for both human and non-human primates (NHPs) \cite{dethier2011brain, dangi2011adaptive, hochberg2012reach, christie2014comparison, dong2023neural, mcmillan2024spike}.

Accurate intra-cortical brain signal decoding is critical for iBCIs. Many existing iBCI decoding approaches use hand-crafted neural activity vector (NAV) features, and linear classifiers/regressors \cite{chen2013sparse, collinger2013high}. Deep learning using artificial neural networks (ANNs) has achieved remarkable successes in computer vision \cite{krizhevsky2012imagenet}, speech recognition \cite{graves2013speech}, natural language processing \cite{devlin2019bert}, and also non-invasive BCIs \cite{drwuPIEEE2023,drwuABAT2024,drwuTTIME2024,drwuNSR2025}. Spiking neural networks (SNNs), which mimic the authentic communication process of neurons in the human brain, are considered as the third generation of neural networks \cite{maass1997networks}. Unlike ANNs that perform calculations using continuous numbers, SNNs utilize sparse and discrete spikes. This manner of calculation enables SNNs to achieve lower energy consumption and higher computational efficiency on neuromorphic hardware \cite{davies2018loihi}, while maintaining comparable performance with ANNs.

SNNs take spiking signals as input, whereas most iBCIs acquire spiking neural signals. Moreover, low energy consumption is very important for iBCIs to avoid damages to the brain. Therefore, SNNs are very promising for iBCI decoding. Unfortunately, there has not been much research in this direction \cite{mcmillan2024spike}.

This paper proposes a spiking neural network for effective and energy-efficient intra-cortical brain signal decoding. We also propose a feature fusion (FF) approach, which integrates the manually extracted NAV features with those extracted by a deep neural network, to further improve the decoding accuracy. Experiments in decoding motor-related intra-cortical brain signals of two rhesus macaques demonstrated that our SNN model achieved higher accuracy than traditional ANNs; more importantly, it was tens or hundreds of times more efficient.

The remainder of this paper is organized as follows. Section \ref{sect:Related_works} introduces related works. Section \ref{sect:Methods} introduces our proposed SNN and FF approaches. Section \ref{sect:Exp} presents the experiment results. Finally, Section \ref{sect:Conclusion} draws conclusions.

\section{Related Works} \label{sect:Related_works}

This section introduces related works on BCIs, SNNs, and SNN for iBCI decoding.

\subsection{Brain Signal Decoding}

ANN-based deep learning has achieved great successes in non-invasive BCIs. Representative approaches include convolutional neural network (CNN) based ShallowConvNet \cite{Schirrmeister2017deeepconv}, DeepConvNet \cite{Schirrmeister2017deeepconv}, and EEGNet \cite{lawhern2018eegnet} , and Transformer based EEGConFormer \cite{song2022eeg} and EEGDeformer \cite{ding2024eeg}.

Existing iBCI decoding algorithms mostly use traditional machine learning models, instead of deep learning. For example, Chen et al. \cite{chen2013sparse} designed a sparse Bayesian linear regression model to decode 3D movements of non-human primates; Zhang et al. \cite{zhang2018decoder} proposed a principal component analysis based algorithm to cope with inter-session discrepancies; Li et al. \cite{li2024thermophysical} introduced a thermodynamic model utilizing both local field potentials and spike signals. In human rehabilitation applications, Collinger et al. \cite{collinger2013high} constructed a neural decoder based on a mathematic model that linearly relates neural firing rates to movement velocities, and calibrated it for several weeks for a 52-year-old patient with tetraplegia. Bradman et al. \cite{brandman2018rapid} used a steady-state Kalman filter and Gaussian process regression on three patients with tetraplegia to achieve rapid closed-loop neural cursor control. Flesher et al. \cite{flesher2021brain} found that it is feasible to mimic known biological control principles in a prosthetic arm controlling system.

\subsection{SNNs}

Spiking neurons, which mimic the biological neurons in the brain by converting continuous signals into spike trains, are the basic elements of SNNs. The Izhikevich model \cite{izhikevich2004model} best approximates the real neurons but suffers from high computational complexity. Simplified neuron models like the quadratic integrate-and-fire (QIF) model \cite{latham2000intrinsic}, the exponential integrate-and-fire (EIF) model \cite{fourcaud2003spike}, the leaky integrate-and-fire (LIF) model \cite{lapicque1907louis}, and the parametric leaky integrate-and-fire (PLIF) model \cite{fang2021incorporating}, are more computationally efficient, and hence used more broadly in SNNs.

Unlikely ANNs, which are usually trained by back-prorogation \cite{Rumelhart1986}, training SNNs through back-propagation is infeasible due to the discrete nature of the spikes, which leads to gradient explosion or vanishing. Generally, there are three main training strategies for SNNs: ANN-to-SNN conversion \cite{diehl2015fast}, spike-timing-dependent plasticity \cite{caporale2008spike}, and surrogate gradient \cite{wu2018spatio}.

\subsection{SNNs for iBCIs}

Several SNNs have been proposed to decode intra-cortical brain signals.

Dethier et al. \cite{dethier2013design} designed a simulated SNN model using the neural engineer framework, a general method for mapping control algorithms into SNNs. Boi et al. \cite{boi2016bidirectional} implemented a spiking network on a compact neuromorphic processor, using the spike-timing-dependent plasticity as the learning method. Their experiments demonstrated the chip's ability to correctly learn decoding tasks. Zheng et al. \cite{zheng2022spiking} utilized an SNN model as data generator, showing that a small amount of training data conforming to neural population dynamics could be generated, thereby enhancing the decoding performance. McMillan et al. \cite{mcmillan2024spike} introduced an SNN-LSTM structure to decode the movement through brain signals, leveraging both the low energy consumption of SNNs and LSTM's ability to capture long-term temporal dependencies.

This paper proposes an SNN for both low energy consumption and high decoding accuracy.

\section{Methods} \label{sect:Methods}

This section introduces our proposed SNN and FF approaches.

\subsection{Spiking Neuron and Surrogate Gradient} \label{spkn}

SNNs transmit information across layers in the form of spike trains through spiking neurons.

For a PLIF neuron \cite{fang2021incorporating} with input $I_t$ and membrane potential $u_t$, its hidden state $h_t$ is updated as:
\begin{align} \label{charge}
	h_t=\beta u_{t} + (1- \beta)I_t,\quad t=1, ..., T
\end{align}
where $T$ is the time step, and $\beta=1-\frac{1}{\tau}$, in which $\tau$ is a learnable positive time constant to ensure $0< \beta < 1$.

The neuron output $O_t$ is determined by the Heaviside function $H$:
\begin{align} \label{fire_eq}
	O_t=H(h_t - V_{\text{th}})=
	\begin{cases}
		1, & \text{if} \enspace h_t \ge V_\text{th}\\
		0, & \text{otherwise}
	\end{cases},
\end{align}
i.e., it fires a spike when $h_t$ exceeds a threshold $V_\text{th}$, and gives no output otherwise.

\begin{figure}[htpb] \centering
\subfigure[]{\label{fig2a} \includegraphics[width=.9\linewidth,clip]{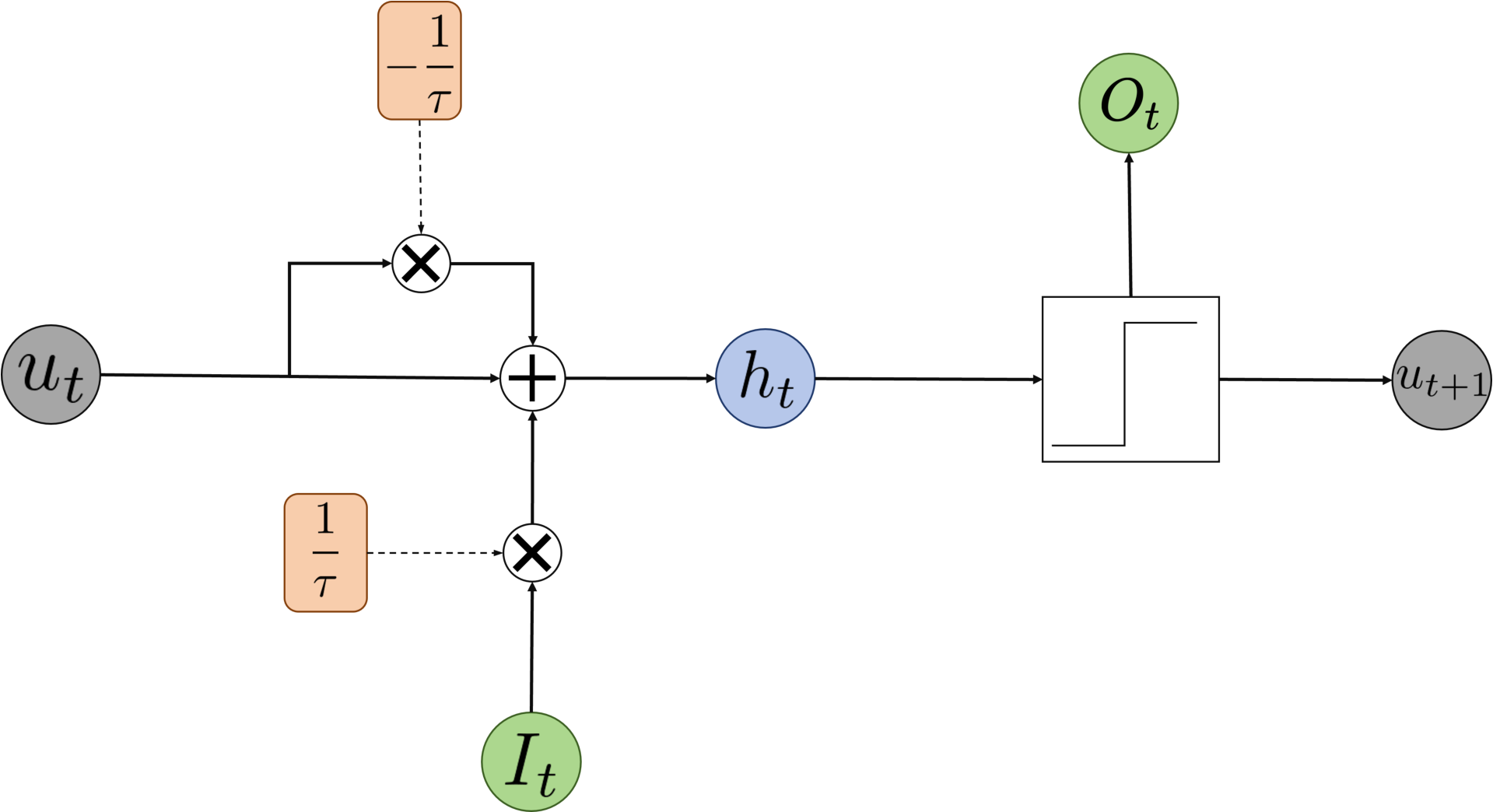}}
\subfigure[]{\label{fig2b} \includegraphics[width=.8\linewidth,clip]{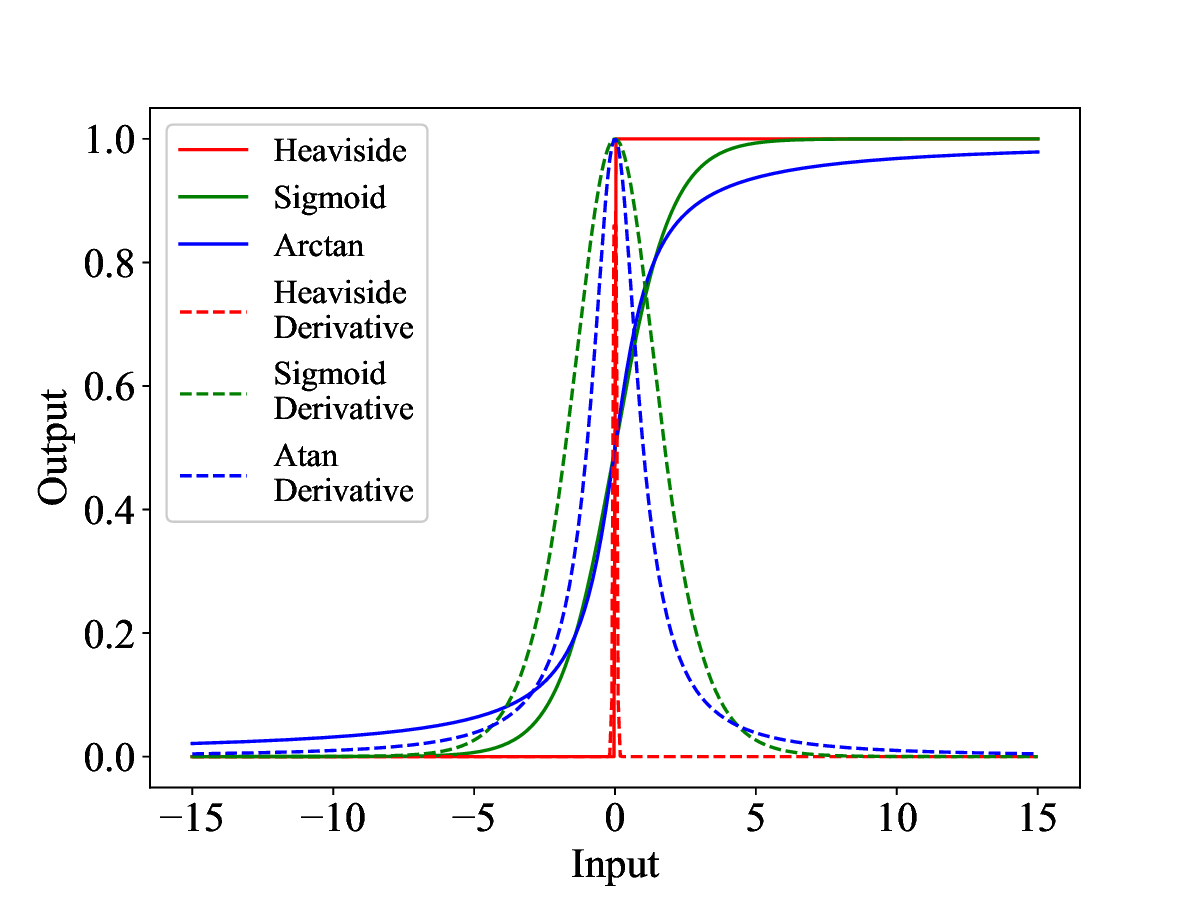}}
\caption{(a) The computational process of a PLIF neuron. (b) The Heaviside function and two surrogate functions, and their derivatives.} \label{fig2}
\end{figure}

If the neuron fires, then its membrane potential resets to $u_\text{reset}$, which by default is set to 0, i.e.,
\begin{align} \centering \label{reset}
 	u_{t+1}=(1-O_t)h_t + O_t u_{\text{reset}}.
\end{align}
Fig. \ref{fig2a} shows the calculation flow of a PLIF neuron.

Computing the gradients for PLIF neuron based SNNs is challenging, because the Heaviside function is non-differentiable: the gradient becomes infinite at the threshold and zero elsewhere, leading to potential gradient vanishing or explosion. Surrogate gradient \cite{neftci2019surrogate} has been proposed to address this problem. It replaces the Heaviside function with a surrogate function (typically sigmoid or arctan) in gradient calculation.

Using the sigmoid function $\sigma(\cdot)$ as an example, the gradient calculation becomes:
\begin{align} \label{surrogate}
 \frac{\partial O_t}{\partial h_t} & = \frac{\partial \sigma(h_t-V_\mathrm{th})} {\partial h_t} \nonumber \\
 		 	&=\sigma(h(t)-V_\mathrm{th})\cdot (1-\sigma(h(t)-V_\mathrm{th})).
\end{align}
This approach, as Fig.~\ref{fig2b} shows, ensures the gradient remains finite and nonzero, approximating the original gradient while enabling more stable training.

\subsection{Our Proposed SNN}\label{SNN}

\begin{figure*}[htpb] \centering
\includegraphics[width=1\textwidth]{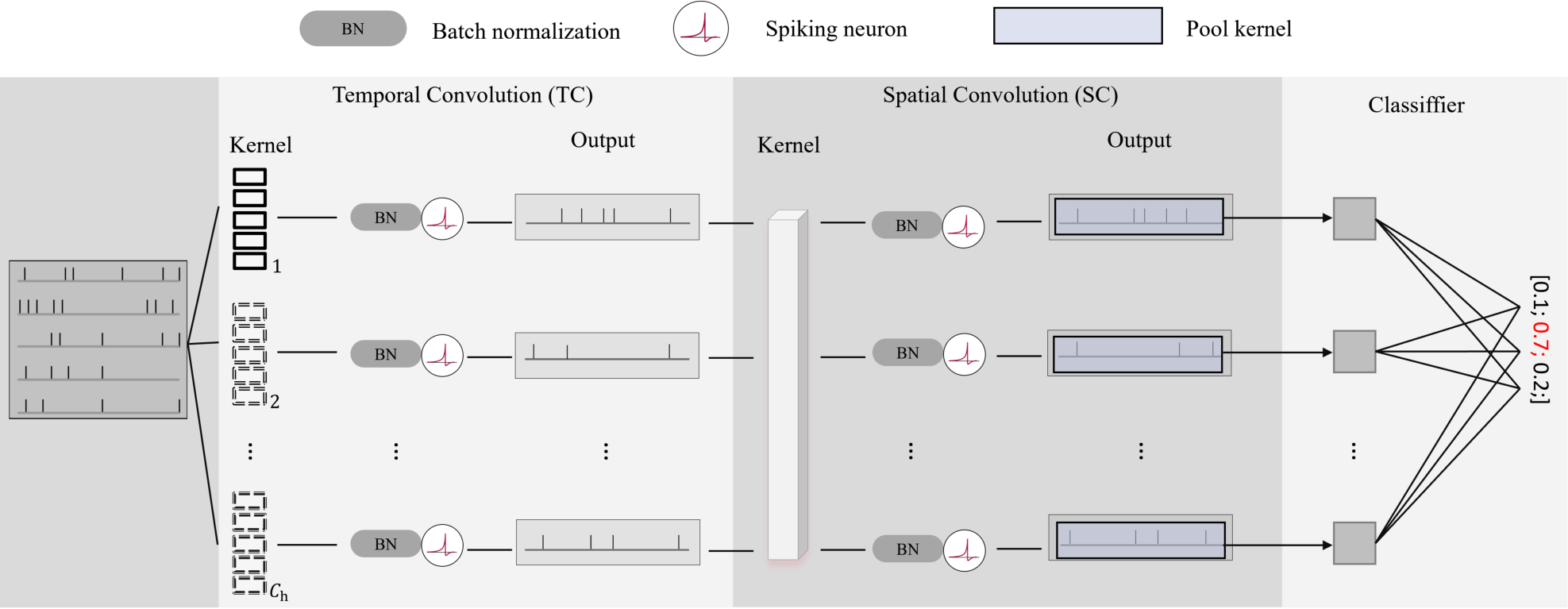}
\caption{Architecture of our proposed SNN. The first convolution layer performs channel-wise temporal convolution, with one convolution kernel for each channel. The second layer is spatial convolution to extract the spatial information across different channels. The third layer is a fully connected layer. In each convolution layer, a batch normalization (BN) layer and a PLIF neuron is employed after the kernel.} \label{fig3}
\end{figure*}

Fig.~\ref{fig3} shows the architecture of our proposed SNN. More specific configurations and the definitions of some variables used in this subsection are given in Table~\ref{tab1}. Its input has the dimensionality $C\times T$, where $C$ is the number of signal channels and $T$ the number of time domain sampling points. $T$ is also used as the time step in the PLIF neurons.

\begin{table*}[htpb] \centering \setlength{\tabcolsep}{4mm}
\caption{The SNN model configuration.} \label{tab1}		
\begin{tabular}{c|ccccccc} \hline
\textbf{Layer} & \textbf{Type} & \textbf{Number of filters} & \textbf{Kernel size} & \textbf{Number of parameters} & \textbf{Output size} \\
\hline
Input & & && & $(C_\text{in},T)$ \\ \hline
Temporal & Conv1D & $C_\text{in}$ & 64 & $C_\text{in} \times C_\text{h}\times 64$ & \\
Convolution & Batch Normalization & $C_\text{h}$ &  &$2C_\text{h}$ & \\
(TC)& Parametric LIF \tnote{*} & &  & $2C_\text{h}$ & $(C_\text{h},T)$ \\ \hline
Spatial & Conv2D & 1 & (64,1) & 64 & \\
Convolution&  Batch Normalization & $C_\text{h}$  & & $2C_\text{h}$ &  \\
(SC)& Parametric LIF \tnote{*} & & & $2C_\text{h}$ & $(C_\text{h},T)$\\ \hline
Average Pooling & & && $(C_\text{h},)$\\ \hline
Classifier & Linear & & & $C\times N_\text{class}$ & $(N_\text{class},)$ \\
\hline
\end{tabular}
\end{table*}

The SNN includes two main components: a feature extractor and a classifier. The feature extractor comprises a temporal convolution (TC) layer and a spatial convolution (SC) layer. The TC layer performs channel-wise 1D convolution and  restructures the input spiking data $X\in \mathbb{R}^{C\times T}$ from $C$ channels to $C_\text{h}$ channels, whereas the SC layer performs standard 2D convolution. Each channel calculated by convolution kernels is followed by a 1D batch normalization layer and a PLIF neuron.

First, the temporal kernel $K_{\text{temp},j}\in \mathbb{R}^{C\times k_\text{temp}}$, where $j=1,2,...,C_\text{h}$ and $k_\text{temp}$ is the kernel length of TC, is applied to the input $X$. This process is:
\begin{align}
		\hat{\mathbf{z}}_j&= \text{cov}(K_{\text{temp},j}, X),
\end{align}
where $\text{cov}$ is the convolution calculation. Denote $\hat{Z}=[\hat{\mathbf{z}}_1;\hat{\mathbf{z}}_2;...;\hat{\mathbf{z}}_{C_\text{h}}]\in \mathbb{R}^{C_\text{h}\times T}$ as the temporal features calculated by the kernels.

The batch normalization layer performs channel-wise normalization using mean $[{\mu}_1;{\mu}_2;...;{\mu}_{C_\text{h}}]$ and variance $[{\sigma}_1; {\sigma}_2;...;{\sigma}_{C_\text{h}}] \in \mathbb{R}^{C_\text{h}}$:
\begin{align}
		\bar{\mathbf{z}}_j &= \frac{\hat{\mathbf{z}}_j-{\mu}_j}{{\sigma}_j}, \\
		\mathbf{z}_j &= \alpha_{j} \cdot \bar{\mathbf{z}}_j + {\xi}_{j},
\end{align}
where $\alpha_{j}$ and $\xi_{j}$ are batch normalization parameters to be tuned. This proecess is denoted as $Z=\text{BN}(\hat{Z})$.

After the batch normalization layer, the temporal feature of the $j$-th channel, $\mathbf{z}_{j}=[z_{1},z_{2},...,z_{T}]$, is sent to the PLIF neuron to calculate the spike encodings by ~\eqref{charge}-\eqref{reset}. For time step $t \in [t_\text{B},t_\text{E}]$, where $t_\text{B}$ is the time the neuron fires the first spike and $t_\text{E}$ is the time the neuron fires the next spike, the hidden state of the neuron is calculated by:
\begin{align}
h_t &= \beta u_t + (1-\beta)z_t \nonumber \\
 &= \beta ^2 u_{t-1} + \beta (1-\beta) z_{t-1} + (1-\beta)z_t \nonumber \\
 &\quad \vdots \nonumber \\
 &= \beta ^ {t-t_\text{B}} (1-\beta) z_{t_\text{B}} + ... + \beta (1-\beta) z_{t-1} + (1-\beta)z_t
\end{align}
The firing process is depicted by \eqref{fire_eq} and \eqref{reset}. The entire calculation flow of the PLIF neuron is denoted as $O=\text{PLIF}(Z)$. Since $\beta$ is learnable, the PLIF neuron output $O$ can adapt to the input $Z$.

The temporal-spatial spike features $F \in \mathbb{R}^{C_\text{h}\times T}$ are next processed by the SC layer in a manner similar to the TC layer, except that the convolution kernel $K_\text{spat}\in \mathbb{R}^{1\times k_\text{spat}}$ ($k_\text{spat}$ is the kernel length of SC) is 2D:
\begin{align}
    F=\text{PLIF}(\text{BN}(\text{cov}(K_\text{spat},O))).
\end{align}

The classifier is a fully connected linear layer with a spiking rate based classification strategy. The spike features $F$ are averaged along the time step dimension to get a feature vector $\mathbf{f} \in \mathbb{R}^{C_\text{h}}$ for classification:
\begin{align}
	\hat y = W_\text{c}\cdot \mathbf{f},\label{eq6}
\end{align}
where $W_\text{c} \in \mathbb{R}^{N_\text{class}\times C_\text{h}}$ is the weight matrix of the classifier, in which $N_\text{class}$ is the number of classes.

\subsection{Feature Fusion}

\begin{figure*}\centering
\includegraphics[width=1\textwidth,clip]{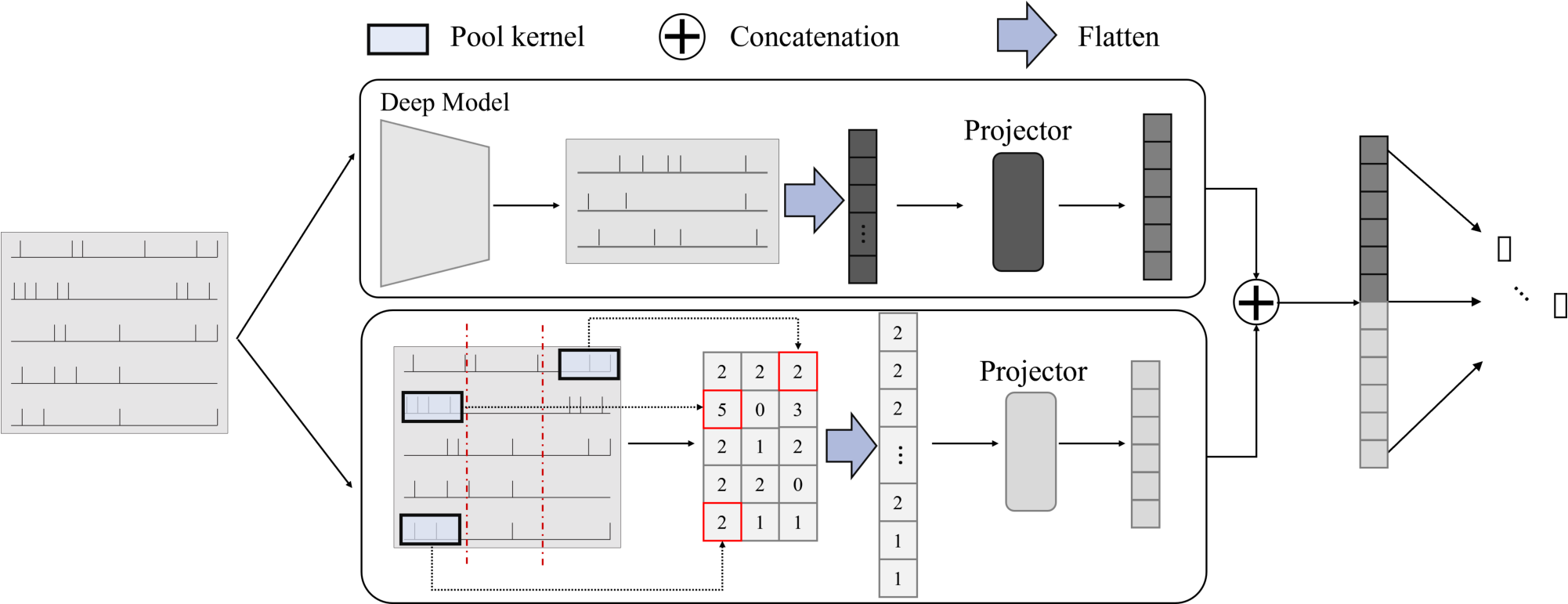}
\caption{Feature fusion. We construct a deep neural network (ANN or SNN) and remove its classifier to extract the deep representations. They and the hand-crafted NAV features are then separately and linearly projected into a hidden space. The two projections are concatenated as input to the linear classifier.} \label{fig4}
\end{figure*}

Prior research \cite{hu2005feature, zhang2018decoder, dong2023neural} has demonstrated the effectiveness of NAV features in decoding movements. Our feature fusion strategy integrates manual NAV features with deep representations extracted by a neural network, as illustrated in Fig.~\ref{fig4}.

To compute the NAV features, for each channel we divide the temporal dimension of the spiking input $X\in \mathbb{R}^{C\times T}$ into $b$ non-overlapping segments of length $T/b$, i.e., $X=[\mathbf{x}_1,...,\mathbf{x}_b]$. We then calculate the spike counts for each segment and obtain the $b$-dimensional NAV features for each channel. The NAV features of different channels are then assembled into feature matrix with dimensionality $C\times b$.

Finally, we project both deep and NAV features into a hidden space and concatenate them. A linear classifier is used to perform the final classification.

\section{Experiments}\label{sect:Exp}

This section performs experiments to demonstrate the effectiveness and efficiency of the proposed SNN and FF approaches. The Python code is available at \url{https://github.com/SongYang008/SNN_iBCIs}.

\subsection{Datasets}

The intra-cortical neural data were collected in our previous research \cite{zhang2018decoder}. The experiment was approved by the Institutional Animal Care and Use Committee at Wuhan University Center for Animal Experiment (No. 2011096) on September 6, 2011.

\begin{figure}[htpb] \centering
\subfigure[]{\centering \label{fig5a} \includegraphics[width=.4\linewidth,clip]{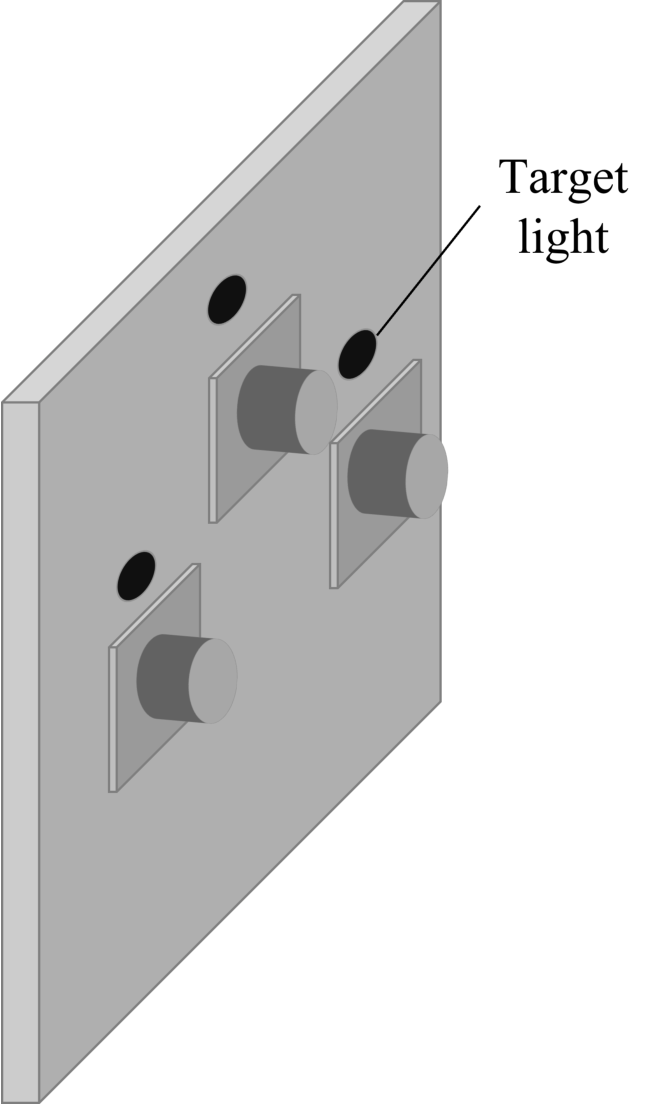}}\hfill
\subfigure[]{\centering \label{fig5b} \includegraphics[width=.48\linewidth,clip]{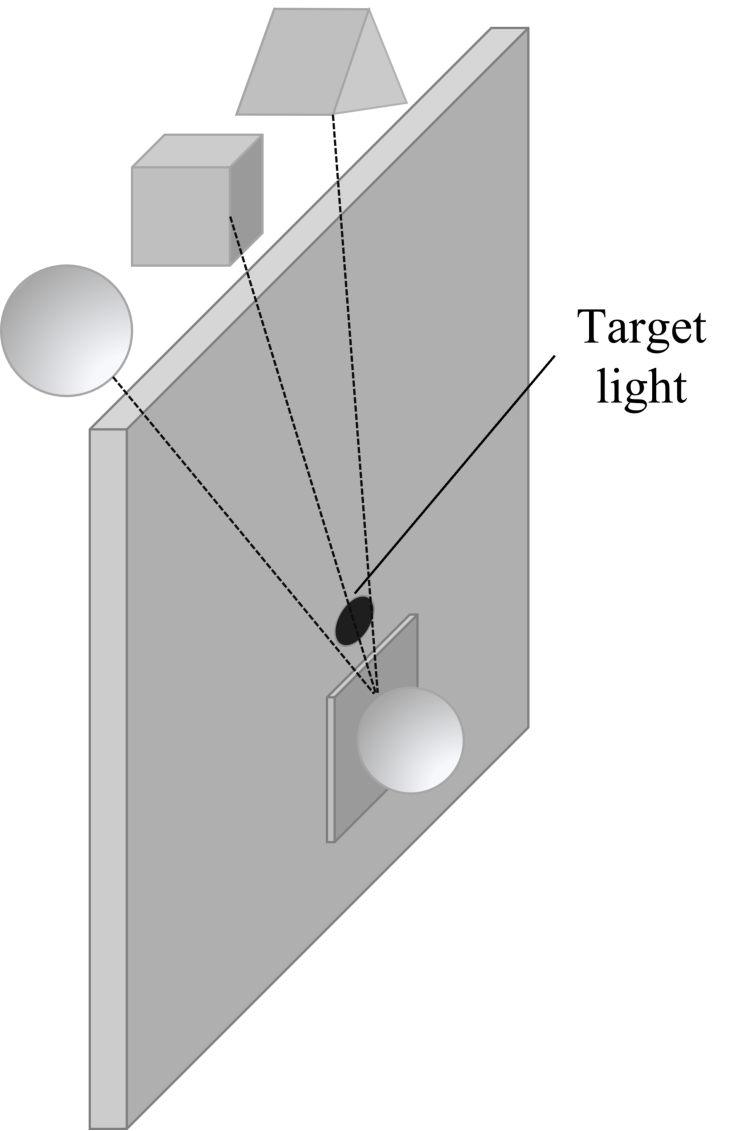}}
\caption{The neural signal recording settings. (a) The reaching task: Three cylindrical objects were placed on a panel, with a target light positioned above each object. In each trial, the monkey performed a reaching movement guided by the illumination of a specific target light. (b) The grasping task: One of three different shapes (cube, triangle, sphere) was presented at the panel's center via a motor, for the monkey to grasp.} \label{fig5}
\end{figure}

Two rhesus macaques (Monkey M and Monkey H) performed distinct motor tasks in the experiments, i.e., Monkey M completed a direction reaching task, and Monkey H executed a shape grasping task (hereafter referred to as reaching task and grasping task), both were three-class classifications. During the experiments, a monkeys sat in front of a panel equipped with objects and target lights, as illustrated in Fig.~\ref{fig5}. For the reaching task, three cylindrical objects were positioned at three different locations on the panel: top, top left, and top right. For the grasping task, a single object was positioned at the panel's center, with its shape varied across trials, cycling through cube, triangle, and sphere.

The neural data were acquired using a multi-electrode sensor consisting of three electrode arrays: two 32-channel Utah arrays and one 16-channel FMA array. They were implanted across three critical cortical regions: the primary motor cortex (M1), primary somatosensory cortex (S1), and posterior parietal cortex (PPC). For the grasping task, only 66 arrays were correctly implanted, resulting in a 66-channel dataset. The reaching task had 80 channels.

As shown in Table \ref{tab2}, the raw spiking signals were initially sampled at 40 kHz. Each trial consisted of 0.3 second of data, corresponding to 12,000 time points. Each trial was then downsampled to 100 time points by taking the sum of every 120 time points. The reaching task had 14 sessions, each with about 750 trials. The grasping task had 12 sessions, each with 400-1000 trials. Different sessions were collected at different time slots of the same day, or in different days.

\begin{table}[htpb] \centering \setlength{\tabcolsep}{1.5mm}
\caption{Characteristics of the two datasets.} \label{tab2}
\begin{tabular}{ccccc}
\hline
\makecell[c]{Task\\ Type} & \makecell[c]{Sampling \\  Frequency} & \makecell[c]{Number of \\ Sampling Points} & \makecell[c]{Number of \\ Channels} & \makecell[c]{Number of \\ Sessions} \\
\hline
Reaching & 40 kHz & 12,000 $\rightarrow$ 100 & 80 & 14\\
Grasping &  40 kHz & 12,000 $\rightarrow$ 100 & 66 & 12\\ \hline
\end{tabular}
\end{table}

\subsection{Experiment Settings and Baselines} \label{exp_set}

We used leave-one-session-out cross-validation in our experiments, i.e., one session was used as the test set and all remaining sessions for training and validation. An 8:2 ratio was used to split the training and validation data. The Adam optimizer with early stopping was used in training, with a maximum of 800 epochs and patience of 100. The initial learning rate was set to 0.001, and the batch size was 256.

The SNN model was implemented using Spikingjelly \cite{fang2023spikingjelly}. The initial $\beta$ of the PLIF neuron was set to $0.5$, and the threshold voltage $V_{\text{th}}$ was 1. The hidden channel size $C_\text{h}$ of our proposed SNN model was $\frac{C}{2}$, i.e., 40 for the reaching task and 33 for the grasping task. In FF, the dimensionality of the hidden space $d$ was set to 128, and the number of dividing groups $b=10$.

Five ANN baselines were considered, including EEGNet, DeepConvNet, ShallowConvNet, EEGConFormer and EEGDeformer. We repeated the experiments five times with different random seeds and report the average accuracy.

\subsection{Classification Performance}

Tables~\ref{tab3} and \ref{tab4} present the classification results, where the best is marked in bold, and the second best by an underlined.

\begin{table*}[t] \centering \setlength{\tabcolsep}{1.5mm}
\caption{Classification performance of different models in the reaching task.}
\label{tab3}
\begin{tabular}{cccccccccccccccc}
\hline
{Session} & 0 & 1 & 2 & 3 & 4 & 5 & 6 & 7 & 8 & 9 & 10 & 11 & 12 & 13 & Avg \\
\hline
ShallowConvNet & 71.70 & 64.16 & 92.23 & 89.28 & 82.83 & 84.10 & 78.80 & 89.79 & 67.53 & 71.50 & 83.11 & 77.42 & \textbf{79.18} & 85.17 & 79.77{$\scriptscriptstyle{\pm}$\tiny{1.41}}\\
DeepConvNet &\textbf{79.29} & 63.62 & 92.50 & 86.62 & 85.41 & 82.07 & {79.36} & 87.49 & 58.92 & 77.80 & 78.47 & 80.28 & {75.82} & 80.74 & 79.17{$\scriptscriptstyle{\pm}$\tiny{1.76}} \\
EEGConFormer & 65.05 &\underline{67.92} & \underline{94.79} & 90.13 & 82.86 & \underline{85.48} & \textbf{81.81} & 88.76 & \underline{71.24} & 81.56 & \underline{85.21} & \underline{82.46} & 73.70 & \textbf{94.09} & \underline{81.79}{$\scriptscriptstyle{\pm}$}\tiny{2.43}\\
EEGDeformer & \underline{78.58} & 56.74 & 93.12 & 90.18 & \underline{86.06} & 84.87 & 75.63 & 90.21 & 67.39 & \underline{82.82} & 82.60 & 80.56 & 71.77 & 87.74 & 80.59$\scriptstyle{\pm}$\tiny{1.68} \\
EEGNet &  73.43 & 66.18 & 94.36 & \underline{93.07} & {89.09} & 84.46 & 73.16 & \underline{90.91} & \textbf{71.80} & {83.94} &{86.74} & 77.52 & 61.53 & \underline{90.90} & 81.22$\scriptstyle{\pm}$\tiny{2.24} \\
\hline
Ours & 77.41 & \textbf{69.66} & \textbf{94.84} & \textbf{93.66} & \textbf{90.07} & \textbf{87.19} & \underline{81.45} & \textbf{92.53} & 67.59 & \textbf{84.81} & \textbf{89.02} & \textbf{85.55} & \underline{76.22} & 90.19 & \textbf{84.25}$\scriptstyle{\pm}$\tiny{1.56}\\
\hline
\end{tabular}
\end{table*}

\begin{table*}[t] \centering
\caption{Classification performance of different models in the grasping task.} \label{tab4}
\begin{tabular}{ccccccccccccccc}
\hline
{Session} & 0 & 1 & 2 & 3 & 4 & 5 & 6 & 7 & 8 & 9 & 10 & 11 & Avg   \\
	\hline
ShallowConvNet &  94.17 & \underline{94.91} & 96.33 & \underline{53.77} & \underline{91.38} & 93.98 & 96.95 & 96.5 & \underline{80.97} & 93.97 & 82.24 & \underline{95.14} & \underline{89.19}$\scriptstyle{\pm}$tiny{1.14}\\
DeepConvNet & 93.46 & 92.53 & 95.94 & 50.18 & 92.28 & 94.61 & 97.41 & 96.76 & 77.87 & 93.1 & 82.58 & 96.02 &  88.56$\scriptstyle{\pm}$\tiny{0.89}\\
EEGConFormer & \textbf{96.02} & \textbf{97.33} & \underline{97.00} & 49.28 & 92.88 & \underline{95.81} & \underline{97.53} & 96.61 & 78.25 & \underline{95.65} & 85.23 & \textbf{97.64} & \underline{89.94}$\scriptstyle{\pm}$\tiny{1.23}\\
EEGDeformer & \underline{95.37} & 95.83 & 97.34 & 53.47 & 93.62 & 95.64 & 98.03 & \underline{96.76} & 74.82 & 95.56 & \underline{86.00} & 97.24 & 89.97$\scriptstyle{\pm}$\tiny{0.81}\\
EEGNet & 94.94 & {96.35} & {97.39} & {57.98} & {94.39} & {95.64} & {97.35} & {97.12} & 78.03 & {95.88} & {87.34} & \underline{97.20} &  {90.80}$\scriptstyle{\pm}$\tiny{0.88}\\
\hline
Ours & {95.68} & 96.29 & \textbf{97.58} & \textbf{62.00} & \textbf{94.84} & \textbf{96.14} & \textbf{97.97} & \textbf{98.27} & \textbf{80.86} & \textbf{96.06} & \textbf{89.02} & {97.51} & \textbf{91.85} $\scriptstyle{\pm}$\tiny{0.73} \\
\hline
\end{tabular}
\end{table*}

For the reaching task, our model outperformed others in most sessions, and achieved the highest average accuracy (84.25\%). EEGConFormer and EEGNet exhibited competitive performance, with average accuracies of 81.79 \% and 81.22 \%, respectively, but their performance was less consistent than our model.

For the grasping task, our model again excelled in multiple sessions, particularly in difficult ones like sessions 3 and 8. It also achieved the highest average accuracy (91.85 \%). EEGNet was the second-best model, with an average accuracy of 90.80 \%. EEGConFormer and EEGDeformer also showed competitive performance in certain sessions.

Overall, the reaching task performance showed lower classification accuracy and larger fluctuations than the grasping task, because the former was more complex. However, in both tasks, our proposed SNN achieved the best average performance. Furthermore, $t$-tests showed that the performance improvements of our proposed SNN model over others were statistically significant ($p<0.01$).

\subsection{Effectiveness of Feature Fusion}

This subsection evaluates the effectiveness of our proposed FF strategy.

First, we investigated the effect of integrating the manually extracted NAV features on both ANN baselines and our proposed SNN model. The results are shown in Table~\ref{tab5}. FF improved the classification performance in most cases for the ANN baselines. Furthermore, it improved the performance of our SNN model in both scenarios: from 84.25\% to 85.19\% in the reaching task, and from 91.78\% to 91.82\% in the grasping task. These results indicated that the NAV features, which spiking rate related, were generally complementary to the deep learning features.

\begin{table}[htpb] \centering
\caption{Classification accuracies (\%) with and without feature fusion.} \label{tab5}
\begin{tabular}{cccccccc}
\hline
\multicolumn{1}{c}{ \multirow{2}{*}{Backbone}} & \multicolumn{2}{c}{Reaching task} & \multicolumn{2}{c}{Grasping task} \\
& w/o FF & w/ FF & w/o FF & w/ FF \\
\hline
ShallowConvNet & 79.77 & 81.20 $\uparrow$ &  89.19 & 90.08 $\uparrow$ \\
DeepConvNet  & 79.17 & 79.36 $\uparrow$ & 88.56 & 90.45 $\uparrow$\\
EEGConFormer & 81.79 & 80.71 \enspace & 89.94 &  90.27 $\uparrow$\\
EEGDeformer &  80.59 & 82.03 $\uparrow$& 89.97 & 90.27 $\uparrow$ \\
EEGNet & 81.22 &  80.73 \enspace & 90.80 & 91.15 $\uparrow$\\
\hline
Ours & \textbf{84.25} & \textbf{85.19} $\uparrow$ & \textbf{91.78} & \textbf{91.82} $\uparrow$ \\
\hline
\end{tabular}
\end{table}

Next, we investigated the effect of the linear feature projectors, i.e., $\text{P}_\text{DF}$ for the deep features and $\text{P}_\text{NAV}$ for the NAV features. Table~\ref{tab6} shows the results. Both projectors improved the classification accuracies in both tasks.

\begin{table}[htpb] \centering
\caption{Classification accuracies (\%) with and without the linear projectors.} \label{tab6}
\begin{tabular}{ccc}
\hline
Model & Reaching task& Grasping task \\
\hline
w/o $\text{P}_\text{DF} \enspace \& \enspace \text{P}_\text{NAV}$ & 84.55 & 91.28\\
w/ $\text{P}_\text{DF}$ & 84.89 & 91.68\\
w/ $\text{P}_\text{NAV}$ & 84.33 & 91.53\\
w/ $\text{P}_\text{DF} \enspace \& \enspace \text{P}_\text{NAV}$  & \textbf{85.19} & \textbf{91.82} \\
\hline
\end{tabular}
\end{table}

\subsection{Ablation Studies}

Ablation studies were performed to investigate the contribution of each module in our proposed SNN model.

\begin{figure}[htpb] \centering
\subfigure[]{ \label{fig6a} \includegraphics[width=.8\linewidth,clip]{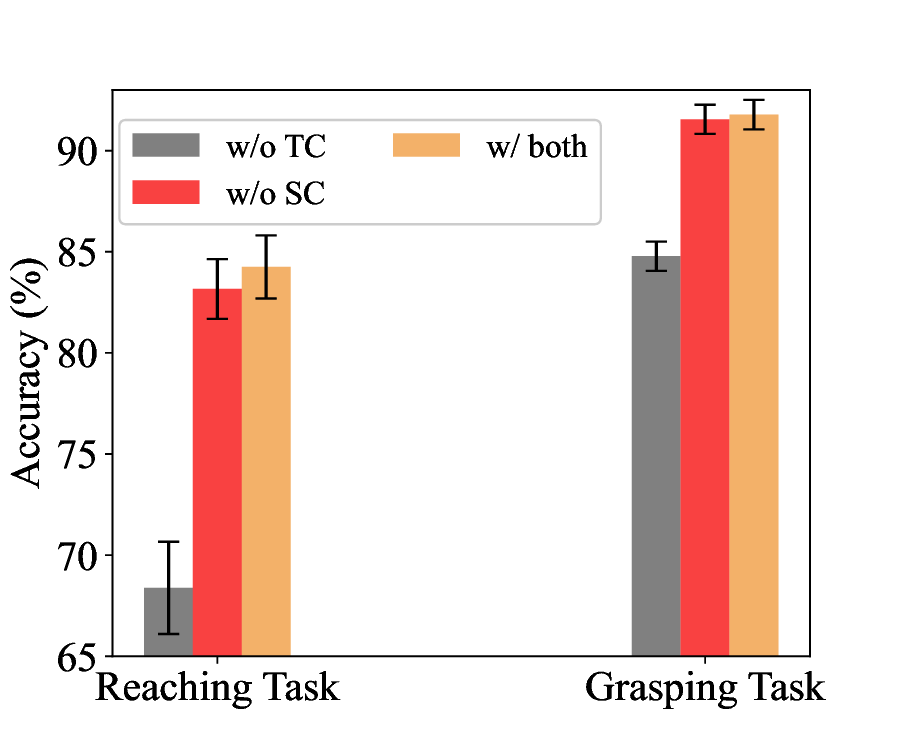}}

\subfigure[]{ \label{fig6b} \includegraphics[width=.8\linewidth,clip]{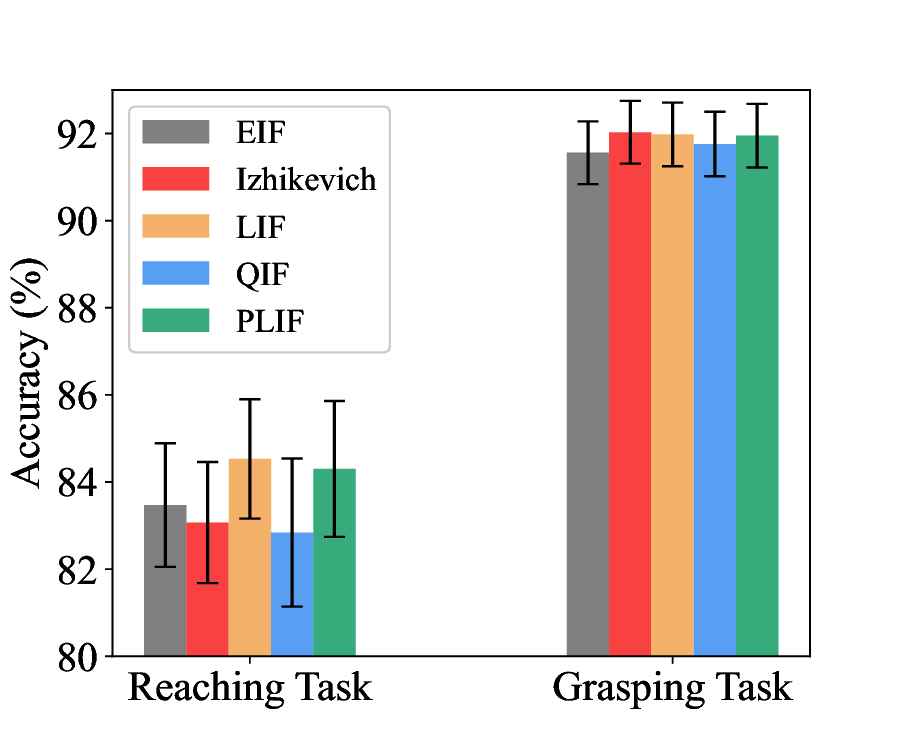}}
\caption{Classification accuracies in ablation studies. (a) The TC and SC layers; and, (b) different spiking neurons.}
\label{fig6}
\end{figure}

\begin{figure}[htpb] \centering
\includegraphics[width=\linewidth,clip]{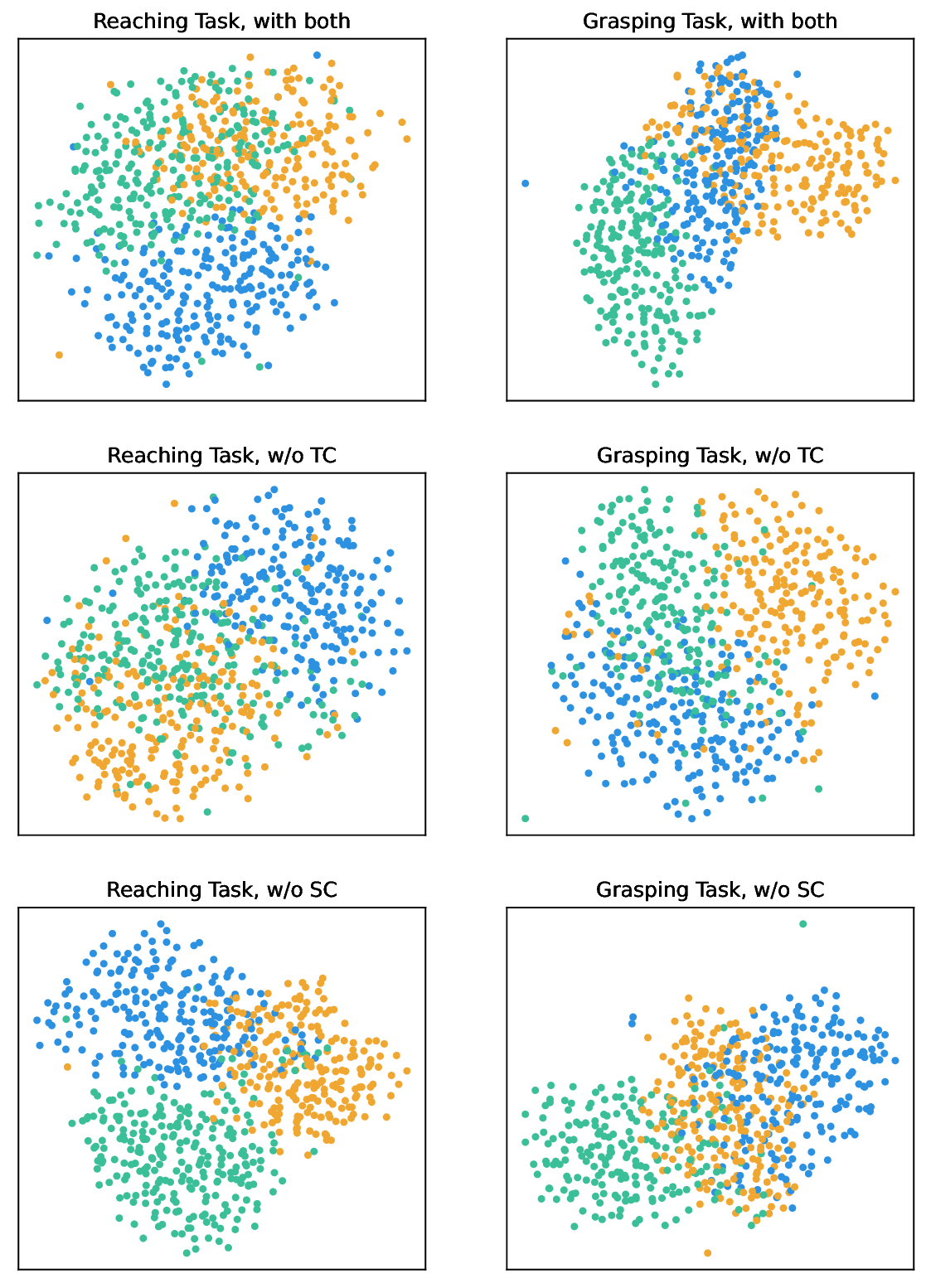}
\caption{$t$-SNE visualization of the first session in the reaching and grasping tasks when removing TC or SC. } \label{fig7}
\end{figure}

First, we examined the role of each convolution layer. Fig.~\ref{fig6a} shows the results. Removing either TC or SC layer degraded the classification accuracy, particularly the TC layer. Fig.~\ref{fig7} further uses $t$-distributed stochastic neighbor embedding ($t$-SNE) \cite{Maaten2008} to visualize the feature distributions after removing the TC or SC layer. It is clear that features from different classes mixed together when the TC layer was removed, explaining why it had significant impact on the decoding performance.

Next, we replaced the PLIF neuron with alternative spiking neurons, including LIF, QIF, EIF and Izhikevich neurons. As shown in Fig.~\ref{fig6b}, PLIF and LIF neurons had the best performance, and they were also the most computationally efficient spiking neurons among the five. Between the two, SNN using the PLIF neuron converged faster in our experiments, so we used it in our model.

\subsection{Energy Consumption Analysis}

As in prior works \cite{yin2021accurate, yao2023attention}, we estimated energy consumption under specific constraints: utilizing 32-bit float-point format data and simulating deployment on a 45-nm technology chip \cite{horowitz20141}. The energy consumption was $E_{\text{MAC}}=4.6$ $p$J for multiplication and accumulation (MAC) operations and $E_{\text{AC}}=0.9$ $p$J for accumulation (AC) operations.

The difference between ANNs and SNNs is that ANNs perform MAC operations, whereas SNNs perform AC operations when taking spiking inputs. For a convolution layer with kernel length $k$, $C_\text{out}$ output channels, and input with dimensionality $C_\text{in} \times T$, where $C_\text{in}$ is the number of input channels, the number of AC operations of an SNN is:
\begin{align}\label{eq:conv_energy}
    N^\text{Cov}_\text{AC} =C_\text{in} \cdot C_\text{out} \cdot T\cdot k \cdot r,
\end{align}
where $r$ is the firing rate of the input spike train.

The number of MAC operations of an ANN is:
\begin{align}
    N^\text{Cov}_\text{MAC}=C_\text{in} \cdot C_\text{out} \cdot T\cdot k.
\end{align}

Assume the input to the batch normalization layer and PLIF neuron has dimensionality $C\times T$. The number of MAC operations is:
\begin{align}
	N^{\text{BN}}_{\text{MAC}}=N^{\text{PLIF}}_{\text{MAC}} = 2 \cdot C \cdot T.
\end{align}

In FF, two linear projectors are adopted. The number of MAC operations is:
\begin{equation}
	N^\text{FF}_\text{MAC}=C \cdot T \cdot d \cdot \frac{1}{b}+C_\text{h} \cdot d.
\end{equation}

The classifier is a fully connected layer. The input dimensionality equals the feature length $d_\text{f}$, and the output dimensionality is the number of classes $N_\text{class}$. The number of MAC operations is:
\begin{align}
	N^{C}_{\text{MAC}}=d_\text{f} \cdot N_\text{class}.
\end{align}

The total energy consumption is hence:
\begin{align} \label{eq:eng}
	E&=E_{\text{MAC}} \cdot N_{\text{MAC}}+E_{\text{AC}}\cdot N_{\text{AC}}.
\end{align}

The energy consumption of an ANN is fixed once the network architecture and input dimensionalities are determined, as the number of MAC and AC operations are determined. In contrast, the energy consumption of an SNN is related to the firing rate, which changes with different inputs. To calculate it more precisely, we calculate the energy consumption of the ANN baselines using an initial model with fixed parameters. For our proposed SNN model, we calculate the average energy consumption based on all samples in the test set.

The results in Table~\ref{tab7} demonstrate the superior energy efficiency of our SNN model. It was at least 35 times more efficient than the ANNs in the reaching task, and at least 59 times more efficient than the ANNs in the grasping task.

\begin{table*}[htpb] \centering \setlength{\tabcolsep}{1.5mm}
\caption{Energy Estimation} \label{tab7}
\begin{threeparttable}
\begin{tabular}{c|cccc|cccc}
\hline
\multicolumn{1}{c|}{\multirow{2}{*}{Model}} & \multicolumn{4}{c|}{Reaching Task} & \multicolumn{4}{c}{Grasping Task} \\
& Params & MAC (M) & AC (K) & Energy Consumption ($\mu$J) & Params & MAC (M) & AC (M) & Energy Consumption ($\mu $J) \\
\hline
ShallowConvNet & 129,243 & 16.16  &  0 &  73.88 (${131} \times$) & 106,843 & 13.33 & 0 &  60.95 (${217} \times$) \\
DeepConvNet & 228,178 &  109.75  & 0 &  504.39 (${898} \times$) & 211,378 & 90.30 & 0 &  415.01 (${1480} \times$) \\
EEGConvFormer & 217,891 &  23.87  & 0 &  109.34 (${194 \times}$) & 222,819 &  29.32 & 0 &  134.55 (${480 \times}$) \\
EEGDeformer &559,803 & 7.55 & 0 & 33.90 ($60\times $) & 502,459 & 7.37 & 0 & 33.90 ($120\times$) \\
EEGNet & 2,531 & 4.47  & 0 &  20.10 (${35} \times$) & 2,307  & 3.67 & 0 & 16.61 (${59} \times$) \\
\hline
Ours & 205,193 & 0.32 & 0.49 & \textbf{0.56} ($\mathbf{1 \times}$)\tnote{*} & 139,729 & 0.27 & 0.17 & \textbf{0.28} ($\mathbf{1 \times}$) \\
\hline
\end{tabular}
\end{threeparttable}
\end{table*}

\subsection{Parameter Sensitivity}

In previous experiments, we set the number of hidden channels $C_\text{h}=\frac{C_\text{in}}{2}$, and the hidden dimensionality $d=128$ in FF. This subsection investigates their impact to the classification performance.

\begin{figure}[htpb] \centering
\subfigure[]{ \label{fig8a} \includegraphics[width=.8\linewidth,clip]{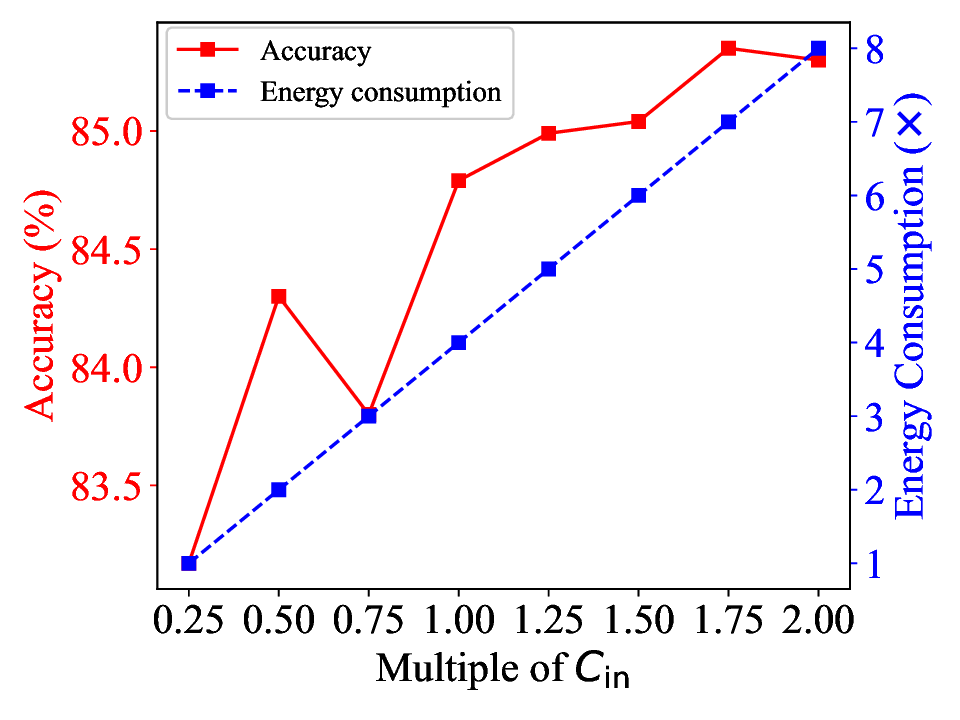}}
\subfigure[]{ \label{fig8b} \includegraphics[width=.8\linewidth,clip]{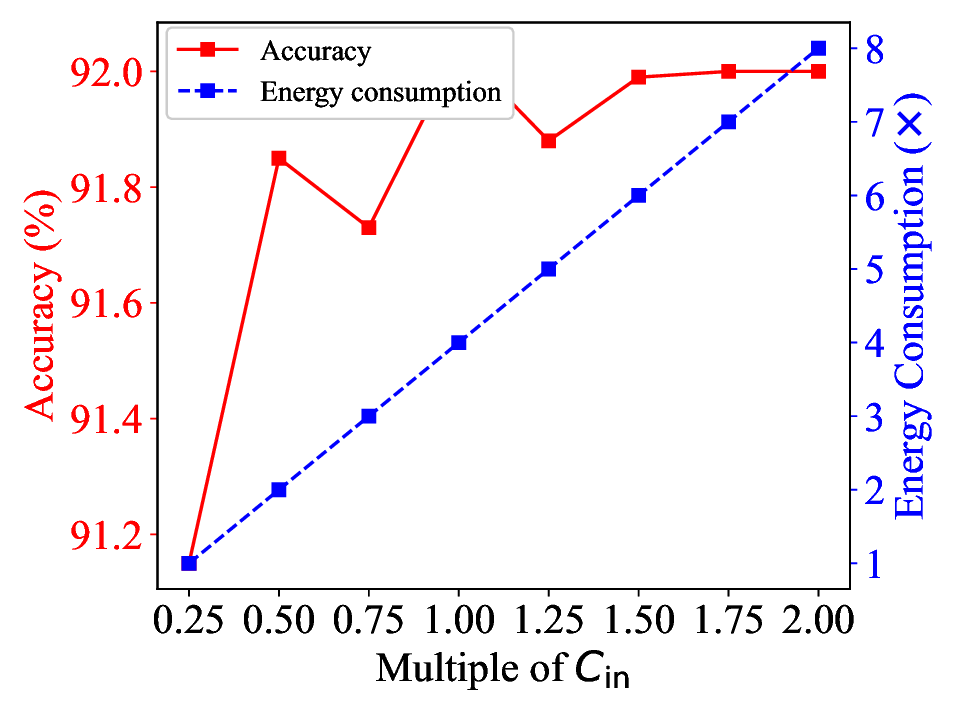}}
\caption{Classification accuracies and normalized energy consumption in the \subref{fig8a} reaching task and \subref{fig8b} grasping task when $C_\text{h}$ varied.} \label{fig8}
\end{figure}

We varied $C_\text{h}$ from $C_\text{in}/4$ to $2C_\text{in}$, i.e., from 20 to 160 in the reaching task and from 16 to 132 in the grasping task. The results are shown in Fig.~\ref{fig8}. Increasing $C_\text{h}$ improved performance, but also led to higher energy consumption. We recommend selecting $C_\text{h}\in[C_\text{in}/2, C_\text{in}]$ to balance the classification performance and efficiency.

\begin{figure}[htpb] 	\centering
	\includegraphics[width=\linewidth,clip]{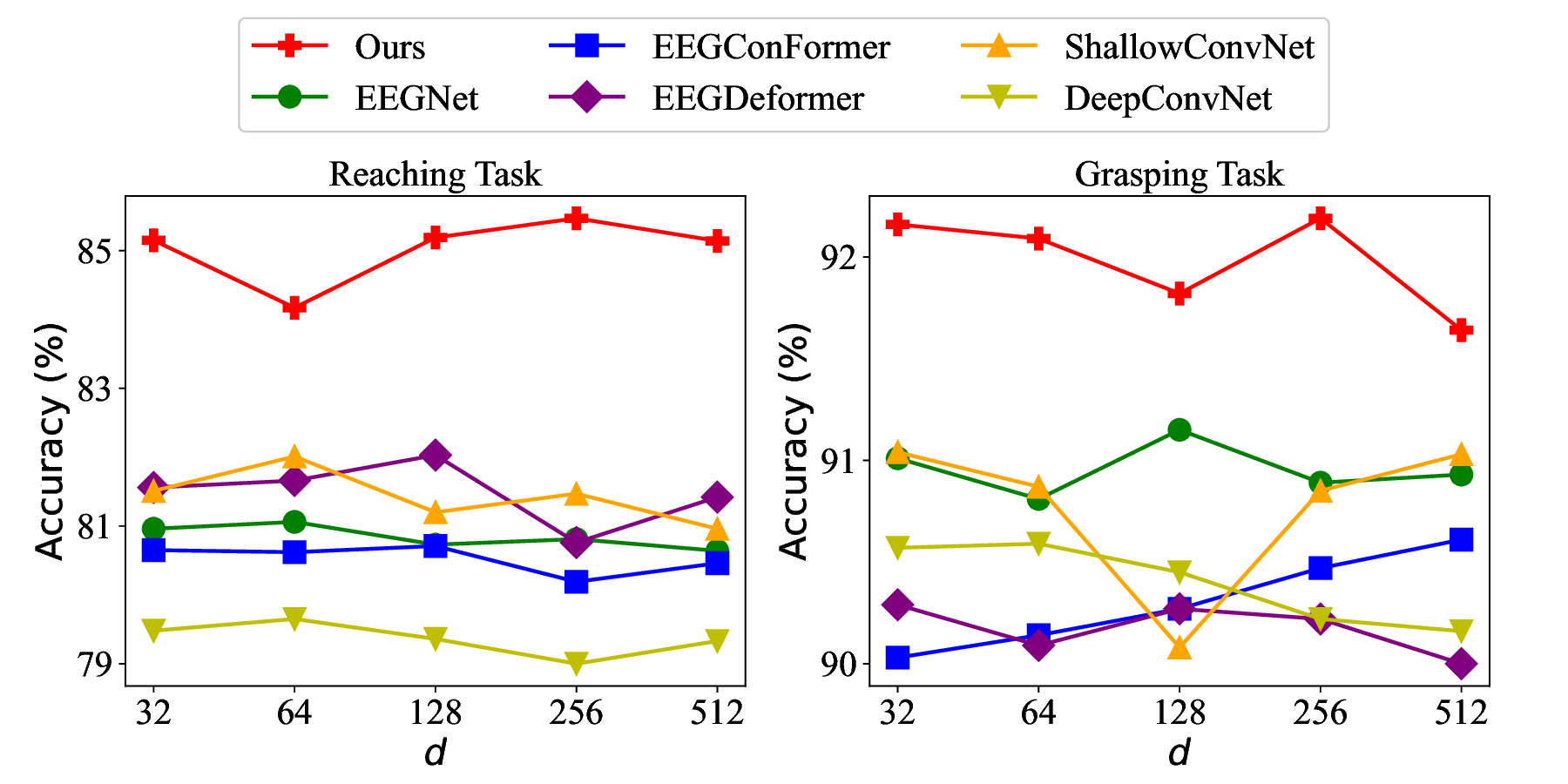}
	\caption{Classification accuracies when the hidden projection dimensionality $d$ in FF varied.} \label{fig9}
\end{figure}

We varied $d$ from 32 to 256. Fig.~\ref{fig9} shows the results. The decoding performance was relatively stable for a wide range of $d$.

\section{Conclusions} \label{sect:Conclusion}

This paper has proposed a spiking neural network architecture for intra-cortical brain signal classification. It consists of a temporal convolution layer and a spatial convolution layer. A batch normalization layer and PLIF spiking neurons are added to each convolution layer. Finally, spike firing rate based features are extracted for classification. We also propose an FF approach, which integrates the manually extracted NAV features with those extracted by a deep neural network, to further improve the decoding accuracy. Experiments on decoding motor-related intra-cortical brain signals of two rhesus macaques demonstrated that our SNN model had higher accuracy than traditional ANNs; more importantly, it was tens or hundreds of times more efficient.



\end{document}